\documentclass[twoside,slac_one]{revtex4}
\usepackage{graphicx}
\usepackage{fancyhdr}
\usepackage{amsmath} 
\usepackage{bm}
\usepackage{amsxtra}
\usepackage{amssymb}
\usepackage{amsthm}
\usepackage{latexsym}
\usepackage{lscape}

\newcommand{\fft}[2]{{\frac{#1}{#2}}}

\def\nn{\nonumber}

\newcommand{\be}{\begin{equation}}
\newcommand{\ee}{\end{equation}}

\def\fft#1#2{\frac{#1}{#2}}

\newcommand{\bea}{\begin{eqnarray}}
\newcommand{\eea}{\end{eqnarray}}

\begin{document}

\title{Strongly-Coupled Quarks and Colorful Black Holes}

%

\author{MD Razikul Islam$^{1}$ and Justin F. V\'azquez-Poritz$^{1,2}$}
\affiliation{$^1$Physics Department, New York City College of Technology, The City University of New York, Brooklyn, NY\\}
\affiliation{$^2$The Graduate School and University Center, The City University of New York, New York, NY}

\begin{abstract}

We use the AdS/CFT correspondence to study the behavior of strongly-coupled quarks in a black hole background. The supergravity background consists of a six-dimensional Schwarzschild-black string AdS soliton, for which the bulk horizon extends from the AdS boundary down to an infra-red floor. By going to higher energy scales, the regime of validity of the classical supergravity background can be extended closer to the singularity than might be expected from the four-dimensional perspective. Small black holes potentially created by the Large Hadron Collider could typically carry color charges inherited from their parton progenitors. The dynamics of quarks near such a black hole depends on the curved spacetime geometry as well as the strong interaction with the color-charged black hole. We study the resulting behavior of quarks and compute the rate at which a quark rotating around the black hole loses energy. We also investigate how the interaction between a quark and an antiquark is altered by the presence of the black hole, which results in a screening length.

\end{abstract}

\maketitle

\thispagestyle{fancy}

\section{Introduction}

Large extra dimensions \cite{large1,large2} could potentially reduce the scale of strong quantum gravity to TeV order. Besides resolving the hierarchy problem, this opens the possibility that small black holes could be produced by the Large Hadron Collider (LHC) \cite{argyres}, and may form within strange stars, which consist entirely of deconfined quark matter \cite{strange}. It is unlikely that such quantum black holes can be described in terms of classical gravity  \cite{quantum}; even the notion of an event horizon is suspect due to potentially large curvature in the vicinity of the would-be horizon. Such a quantum black hole could carry electric charge, color charge and angular momentum inherited from the partons that created it. BlackMax is a black hole event generator that can account for all gauge charges and is being used to put constraints on mini black hole production at the LHC \cite{blackmax1,blackmax2}.

The black hole would interact with nearby quarks via gravity as well as color and electrical interactions. However, there are two potential roadblocks for describing such a system. Firstly, the semi-classical approach to gravity is unreliable when the quarks are close to the black hole. Secondly, the color interaction might take place at  strong coupling, in which case conventional quantum field theoretic methods are ill-equipped to describe dynamical processes. We will demonstrate that the AdS/CFT correspondence \cite{agmoo} may be used to evade both of the difficulties. We will consider a black hole with only a small amount of color charge so that we can neglect the effect of the color charge itself on the spacetime geometry. We are also making the working assumption that the AdS/CFT correspondence can be applied even though there is a singularity extending out to the boundary.

The dynamics of quarks in strongly-coupled plasmas have been studied extensively by considering open strings near a black brane that is extended along the spacetime directions of the field theory; see for example \cite{herzog}. In contrast, a strongly-coupled system of quarks near a black hole corresponds to a supergravity background that contains a black string extending along the bulk radial direction, various possibilities of which have been discussed in \cite{hubeny}. As one might expect, for both of these scenarios there are a number of similarities in the behavior of the quarks.

\section{Some Preliminaries}

\subsection{The Supergravity Background}

We consider a solution to six-dimensional Einstein gravity which is closely related to the AdS soliton \cite{horowitz1}, whose metric is
\be\label{metric5}
ds_6^2=\fft{r^2}{R^2}\left( h dy^2+ds_4^2\right)+\fft{R^2}{r^2} h^{-1}\ dr^2,\qquad h=1-\fft{b^5}{r^5},
\ee
where the $y$ direction is circular and corresponds to a large extra dimension. Although we consider the simplest scenario in which there is a single extra dimension, this is readily generalized to an arbitrary number of flat extra dimensions. Also, evidence for the existence of a class of generalizations of the AdS soliton with the circle $S^1$ replaced by a sphere $S^m$ was presented in \cite{soliton-sphere}. For the AdS soliton, $ds_4^2$ is the Minkowski metric. A Schwarzschild-black string AdS soliton \cite{hubeny} can be constructed by simply replacing $ds_4^2$ with the Schwarzschild metric
\be\label{metric4}
ds_4^2=-f dt^2+f^{-1} d\rho^2+\rho^2 \left( d\theta^2+\sin^2\theta\ d\phi^2\right),\qquad f=1-\fft{2M}{\rho}.
\ee
There is a singularity at $\rho=0$ which runs  all along the $r$ direction and is shielded by a horizon at $\rho=2M$. For vanishing $b$ there is a nakedly singular surface at $r=0$. This is avoided by restricting the range of the radial coordinate to be $r\ge b>0$. $r=b$ corresponds to an ``infra-red floor'' where the $S^1$ smoothly caps off like a cigar geometry, provided that the coordinate $y$ has a periodicity of $\Delta y=4\pi R^2/(5b)$. Since the geometry is asymptotically locally AdS, the Schwarzschild-black string AdS soliton is conjectured to describe a strongly-coupled field theory near a black hole, with one extra dimension.

Let us now examine the regime of validity of the classical supergravity background. The Kretschmann scalar is given by
\be
K=\fft{12}{R^4} \left( 5+\fft{20b^{10}}{r^{10}}+\fft{4M^2 R^8}{\rho^6 r^4}\right),
\ee
which is small as long as
\be
R\ll 1,\qquad \rho\gg M \left( \fft{R}{r}\right)^{2/3}.
\ee
Compare this to the perspective of the four-dimensional field theory, for which the black hole background has small curvature for $\rho\gg M$. Thus, at an energy scale $r>R$, the AdS/CFT correspondence enables us to probe regions closer to the black hole than could be done by using the semi-classical approach directly with the four-dimensional theory.

The above six-dimensional background can be embedded within massive type IIA theory \cite{massiveIIA} in ten dimensions by using the Kaluza-Klein reduction given in \cite{6dbh}. The ten-dimensional metric is
\be\label{10dmetric}
ds_{10}^2=s^{-1/3} \left( ds_6^2+g^{-2} (d\alpha^2+c^2 d\Omega_3^2)\right),
\ee
where $s=\sin\alpha$, $c=\cos\alpha$ and the gauge coupling constant $g$ is related to the mass parameter $m$ of the massive Type IIA theory by $m=\sqrt{2}g/3$. This is a warped product of the six-dimensional space and a hemisphere of $S^4$. The warp factor is singular at the boundary at $\alpha=0$ and the string coupling diverges there since $e^{\Phi}=s^{-5/6}$.

\subsection{Embedding Flavor D8-Branes}
 
The DBI action for a D8-brane is given by
\be
S_{D8}=-T_8\int d^9\sigma\ e^{-\Phi}\sqrt{-\mbox{det}(g_{ab})},
\ee
where $g_{ab}=G_{MN}\partial_a X^M \partial_b X^N$ is the induced metric on the worldvolume of the D8-brane with worldsheet coordinates $\sigma^a$. $G_{MN}$ is the ten-dimensional metric with $\partial_a=\partial /\partial \sigma^a$. We can consider an embedding in which the D8-branes extend in four-dimensional spacetime, are wrapped on the internal 4-sphere, and form a curve for slicings of constant $\rho$ described by the function $r(y,\rho,\alpha)$. 
For the case $r=r(y)$, the first integral of the equation of motion reduces to
\be
(\partial_yr)^2=\fft{r^4h^2}{r_0^{10} h_0 R^4} \left( r^{10}h-r_0^{10}h_0\right),
\ee
where the D8-branes curve down to the minimal radius of $r=r_0$, and $h_0=h(r_0)$. While there is a ``trivial'' solution with $u_0=0$ which represents disjoint D8-branes, the curved solution is the energetically favored one.

\section{Single-Quark String Configurations}

The dynamics of a classical string are governed by the Nambu-Goto action
\be
S=-\fft{1}{2\pi\alpha^{\prime}} \int d^2\sigma \sqrt{-\mbox{det}(g_{ab})}\,,
\ee
where $g_{ab}=G_{MN}\partial_a X^M \partial_b X^N$ is the induced metric on the string worldsheet. We will consider strings which lie at a point away from $\alpha=0$ on the 4-sphere that was used to lift the six-dimensional Schwarzschild-black string AdS soliton to ten dimensions. Thus, $X^M$ run over the six coordinates ${t,\rho,\theta,\phi,y,r}$. Since $\alpha^{\prime}$ can be canceled out in the action anyway by using the radial coordinate $u=r/\alpha^{\prime}$, we set $\alpha^{\prime}=1$

\begin{figure}[h]
\centering
\includegraphics[width=140mm]{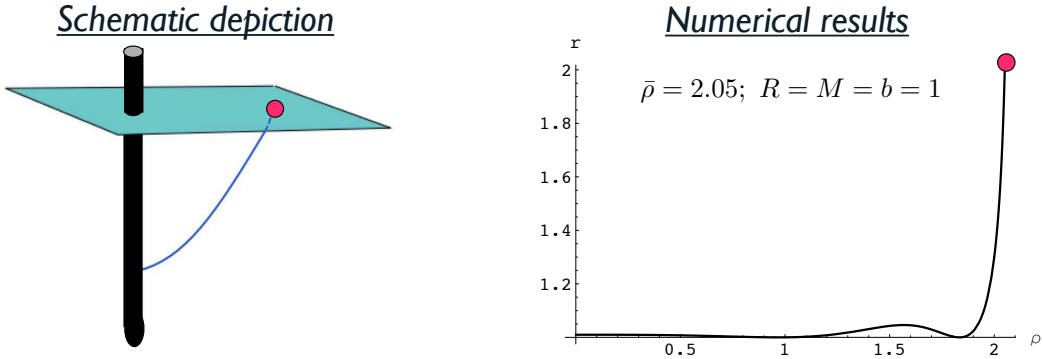}
\caption{The picture on the left is a schematic depiction of the string configuration corresponding to a quark interacting with a color-charged black hole. The corresponding numerical plot is shown on the right.} 
\label{fig2}
\end{figure}

For a single-quark string configuration, an endpoint lies on a probe D-brane and the string goes through the horizon at $\rho=2M$ and approaches the singularity at $\rho=0$, as is schematically depicted in the left picture of Figure \ref{fig2}. Due to the spherical symmetry of the four-dimensional spacetime, a static string configuration lives within the $r-\rho$ plane. However, we will consider a string that is steadily moving in the $\phi$, for which we will allow for the possibility that the string spirals around the $\phi$ direction as well. This string configuration can be described by the worldsheet embedding
\be\label{embedding1}
t=\tau,\qquad \rho=\sigma,\qquad \theta=\fft{\pi}{2},\qquad \phi= \phi(\sigma)+\omega\tau,\qquad r=r(\sigma),
\ee
with the boundary conditions
\be
0\le \tau\le T,\qquad 0\le\sigma\le\bar\rho,\qquad r(\bar\rho)=\bar r,
\ee
where the endpoint of the string is located at $r=\bar r$ and $\rho=\bar \rho$. $\omega$ and $v=\bar\rho\omega$ are respectively the angular velocity and velocity of the string endpoint measured by an observer far from the black hole. On the other hand, the proper velocity $V$ of the string endpoint is given by
$V=v f^{-1/2}|_{\rho=\bar\rho}$. Thus, in order for the string to be timelike, we must have $V<1$ which translates into
\be
v<\sqrt{1-\fft{2M}{\bar\rho}}.
\ee
With the above worldsheet embedding, the action becomes
\be\label{S1}
S=-\fft{T}{2\pi} \int d\rho \sqrt{\left( f-\omega^2\rho^2\right)\left(\fft{r^4}{R^4}f^{-1}+h^{-1}r^{\prime 2}\right)+\fft{r^4}{R^4} f\rho^2\phi^{\prime 2}},
\ee
where $^{\prime}\equiv\partial/\partial\rho$. The equations of motion are given by
\bea\label{eom3}
\fft{d}{d\rho} \left[ A (f-\omega^2\rho^2)h^{-1} r^{\prime}\right] &=& \fft{A}{R^4} \left[ (f-\omega^2\rho^2) \left( 2r^3f^{-1}-\fft{5R^4b^5}{2h^2r^6} r^{\prime 2}\right)+2\rho^2 r^3f\phi^{\prime 2}\right],\nn\\
\fft{d}{d\rho} \left[ Ar^4f\rho^2\phi^{\prime}\right] &=& 0,
\eea
where
\be\label{A}
A^{-1}=\sqrt{(f-\omega^2\rho^2)(r^4f^{-1}+R^4h^{-1}r^{\prime 2})+\rho^2r^4f\phi^{\prime 2}}.
\ee
The second equation in (\ref{eom3}) can be integrated once to give
\be\label{phi}
\phi^{\prime}=\fft{c}{Ar^4f\rho^2},
\ee
where $c$ is an integration constant that parameterizes how much the string spirals around the $\phi$ direction. In particular, $c=0$ corresponds to a string that lives solely in the $r-\rho$ plane. 

The canonical momentum densities associated to the string are given by
\bea
\pi_{\mu}^0 &=& -h_{\mu\nu}\ \fft{(\dot X\cdot X^{\prime})(X^{\nu})^{\prime}-(X^{\prime})^2 (\dot X^{\nu})}{2\pi \sqrt{-G}},\nn\\
\pi_{\mu}^1 &=& -h_{\mu\nu}\ \fft{(\dot X\cdot X^{\prime})(\dot X^{\nu})-(\dot X)^2 (X^{\nu})^{\prime}}{2\pi \sqrt{-G}}.
\eea
The total energy and momentum in the $\phi$ direction of the string are
\be
E=-\int d\sigma\ \pi_t^0,\qquad p_{\phi}=\int d\sigma\ \pi_x^0.
\ee
The string will gain energy through its endpoint at a rate of
\be
\pi_t^1=\fft{\omega c}{2\pi R^2}.
\ee
The string exerts a force on its endpoint, which must be countered by an external force. For a purely radial string, the external force must be radially outwards. If the string spirals around the $\phi$ direction, then we also need to apply a force in the $\phi$ direction given by
\be
\pi_{\phi}^1=\fft{c}{2\pi R^2}.
\ee
From the four-dimensional gravity perspective this is rather surprising, given that the quark is orbiting the black hole at a constant speed and the black hole is not rotating. However, this is a result of the strong interaction between the quark and the black hole.

From (\ref{A}) and (\ref{phi}), we find
\bea
-G &=& \fft{1}{R^4 A^2} = \fft{\rho^2 r^4}{R^4} \left(\fft{f-\omega^2\rho^2}{r^4f\rho^2-c^2}\right) (r^4+R^4fh^{-1}r^{\prime 2}),\nn\\
\phi^{\prime 2} &=& \fft{c^2}{r^4f^2\rho^2} \left( \fft{f-\omega^2\rho^2}{r^4f\rho^2-c^2}\right) (r^4+R^4fh^{-1}r^{\prime 2}).
\eea
In order to have a well-defined string and avoid parts of it moving faster than the local speed of light, $-G$ must be positive definite all along the string. For nonzero $c$, this implies that $\phi^{\prime 2}>0$. Thus, the string spirals around the $\phi$ direction as it approaches the black hole, without reversing direction. 

In order for $-G>0$, any real roots of the numerator of $-G$ that lie along the string must also be roots of its denominator. Consider the function
\be\label{F}
F\equiv f-\omega^2\rho^2.
\ee
If there is a point $\rho=\rho_{crit}\le \bar\rho$ on the string such that $F(\rho_{crit})=0$, then this must specify the value of $c$ via
\be\label{c}
r_{crit}^4 f_{crit} \rho_{crit}^2-c^2=0,
\ee
where $r_{crit}\equiv r(\rho_{crit})$ and $f_{crit}\equiv f(\rho_{crit})$. Since $r_{crit}$ depends on $c$, the specified value of $c$ must be found numerically. On the other hand, if $\rho_{crit}>\bar\rho$, then $c$ is unspecified. In this case, given initial conditions with a nonvanishing value of $c$, the string will presumably evolve to the minimum-energy configuration with $c=0$.

Let us first consider a static string with $\omega=0$. If the endpoint lies outside of the horizon ($\bar\rho>2M$) then, in order to have a well-defined string outside and on the horizon, we must have $c=0$. For the numerical integration, we take the boundary conditions to be at an intermediary point on the string $\rho=\rho_0$ and $r(\rho_0)=r_0$. Taking $\rho_0=2$ corresponds to a string which extends through the horizon, although the string endpoint is of course only sensitive to the portion of the string that lies outside of the horizon. In order to do the numerical integration, we have taken $\rho_0=2+10^{-8}$ for the region outside the horizon and then matched this with the region inside the horizon for $\rho_0=2-10^{-8}$. The numerical result for $\bar\rho=2.05$ and $R=M=b=1$ is given in the right plot of Figure \ref{fig2}. Note that this plot is no longer reliable as one gets too close to $\rho=0$. The undulations in the $r$ direction is a feature associated with nonvanishing $b$ parameter of the AdS soliton, which can be demonstrated by taking $b=0$ and finding that the curve of the string is monotonic in $r$. 

Now we will consider steadily-rotating strings. The constraints on $c$ can be determined from the roots of (\ref{F}), which is a cubic function of $\rho$. The discriminant is
\be
\Delta=4\omega^2 (1-27M^2\omega^2).
\ee
For $\omega> \omega_1\equiv 1/(\sqrt{27}M)$, $\Delta<0$ and there is one real root and two nonreal complex conjugate roots. Since $F(\rho\rightarrow -\infty)\rightarrow -\infty$ and $F(\rho\rightarrow 0^-)\rightarrow +\infty$, there is always a real negative root. Thus, there are no physical roots of $F$ for $\omega>\omega_1$. Then in order for the denominator of $-G$ to never vanish, we must have the lower bound $c^2\ge c_{min}^2$, where $c_{min}$ satisfies (\ref{c}). Note that $c_{min}\neq 0$ if the string endpoint is on or outside of the horizon, while $c_{min}=0$ if the endpoint lies within. In other words, for the case in which the string endpoint is outside of the horizon and moving at a high speed, there is a minimal amount with which the string must spiral around the $\phi$ direction.

For $\omega\le \omega_1$, (\ref{F}) has three real roots, which are all distinct for $\omega<\omega_1$ ($\Delta>0$) and two of which coincide for $\omega=\omega_1$ ($\Delta=0$). In particular, in addition to the real negative root, there are two real positive roots which lie outside of the  horizon for $\omega>0$. If one of them lies within the region $2M<\rho_{crit}\le \bar \rho$, then this specifies the value of $c$ via (\ref{c}). From (\ref{F}), we see that this occurs for $\omega\le \omega_2\equiv\sqrt{1-2M/\bar\rho}$. On the other hand, for $\omega>\omega_2$, then we must have the lower bound $c^2>c_{min}^2$. Note that the threshold velocity vanishes for quarks at the horizon and approaches the speed of light for quarks far from the black hole. 

To summarize, for a string endpoint moving with angular velocity $\omega$ whose endpoint lies outside of the horizon, the value of $c$ is specified for $\omega\le \mbox{Min}(\omega_1,\omega_2)$ while we have the lower bound $c^2>c_{min}^2$ for $\omega> \mbox{Min}(\omega_1,\omega_2)$, where $\mbox{Min}(\omega_1,\omega_2)$ is the minimum quantity between $\omega_1$ and $\omega_2$. On the other hand, if the endpoint lies within the horizon, then the value of $c$ can be arbitrary.

\begin{figure}[h]
\centering
\includegraphics[width=62mm]{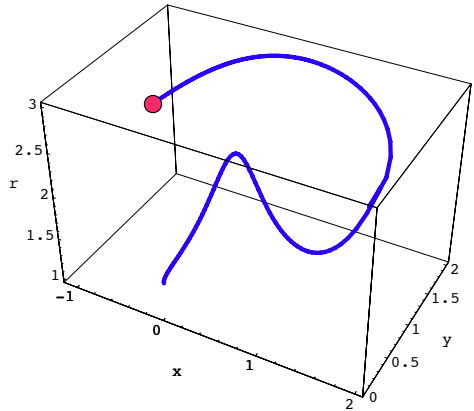}
\caption{A steadily-rotating, spiralling string for $\omega=0.5$, $b=c=1$ and $\bar\rho=2.02$.} 
\label{fig10}
\end{figure}

Figure \ref{fig10} shows a steadily-rotating, spiralling string whose endpoint lies outside of the horizon at $\bar\rho=2.02$ and is moving at $\omega=0.5$. We have taken the parameters $R=M=b=c=1$. Since $\omega>\mbox{Min}(\omega_1,\omega_2)\approx 0.1$, $c$ must satisfy $c^2>c_{min}^2$, which can be confirmed numerically. We have matched the solution outside of the horizon with $\rho_0=2.0005$ and $r_0=3$ to the solution within the horizon with $\rho_0=1.991$ and $r_0=3$. Spiralling strings near a black brane extended along the spacetime directions of the field theory have been considered, for example, in \cite{spiralling}.

\begin{figure}[h]
\centering
\includegraphics[width=72mm]{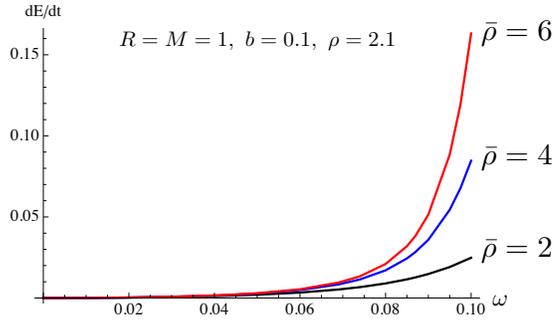}
\caption{Rate of energy loss versus angular velocity.} 
\label{fig11}
\end{figure}

As a quark circles around the black hole it loses energy, which is analogous to the energy loss of a quark moving in a strongly-coupled plasma \cite{herzog}. We have numerically computed the rate of energy loss as a function of the angular velocity $\omega$ for various values of the mass parameter $\bar\rho$, as shown in Figure \ref{fig11}. 
We used the shooting method in order to find the value of $c$ for a given $\omega$ that enforces the boundary condition $r(\bar\rho)=\bar r$. As can be seen, the rate of energy loss increases with angular velocity, as well as with the mass parameter. We have found that the results can be rather well fitted to the following formula:
\be
\fft{dE}{dt}\approx 0.008\omega \exp\left[ (20.65+16.45\bar\rho-0.25\bar\rho^2)\omega-0.89\bar\rho\right].
\ee
%

\section{Static Quark-Antiquark String Configurations}

\subsection{Separation Along the Radial Direction}

\begin{figure}[h]
\centering
\includegraphics[width=71mm]{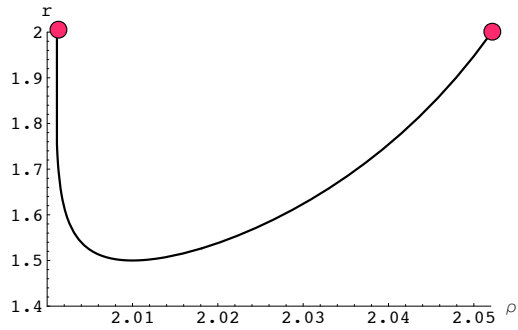}
\caption{A quark-antiquark string configuration with endpoints separated in the radial direction.} 
\label{fig1}
\end{figure}

A static quark-antiquark string with endpoints separated in the $\rho$ direction is described by the worldsheet embedding (\ref{embedding1}) with $\omega=0$, except now with the boundary conditions 
\be
0\le \tau\le T,\qquad \bar\rho_1\le\sigma\le\bar\rho_2,\qquad r(\bar\rho_1)=r(\bar\rho_2)=\bar r.
\ee
We will not consider light-heavy quark-antiquark pairs here, for which $r(\bar\rho_1)\ne r(\bar\rho_2)$. At the turning point, the equation of motion gives
\be\label{turning-point}
\partial_{\rho}^2r(\rho_0)=\fft{2r_0^3 h_0}{R^4 f_0},
\ee
where $f_0\equiv f(r_0)$ and $h_0\equiv h(r_0)$. For $r_0>b$, this is positive outside of the black hole which implies that outside of the black hole there are strings which descend below the probe brane and then turn back up. A numerical plot of a quark-antiquark string configuration is shown in Figure \ref{fig1}. The value of the turning point $r_0$ is chosen such that the string endpoint separation at $\bar r=2$ is kept fixed at approximately $0.05$, and we have taken the $\rho$ turning point to be at $\rho_0=2.01$. It can be seen that the midsection of the string is pulled towards the horizon, which is associated with a modification of the quark-antiquark potential due to the presence of the black hole.

\subsection{Separation Along the Tangential Direction}

A quark-antiquark pair separated along the $\phi$ direction can be described by the worldsheet embedding
\be
t=\tau,\qquad \rho=\rho(\sigma),\qquad \theta=\fft{\pi}{2},\qquad \phi=\sigma,\qquad r=r(\sigma),
\ee
with the boundary conditions
\be
0\le \tau\le T,\quad -\bar\phi_1\le\sigma\le \bar\phi_2,\quad r(\bar\phi_1)=r(\bar\phi_2) =\bar r,\quad \rho(\bar\phi_1)=\rho(\bar\phi_2)=\bar\rho.
\ee
Then the action becomes
\be
S=-\fft{T}{2\pi} \int d\phi \sqrt{\rho^2 \fft{r^4}{R^4} f+f h^{-1} r^{\prime 2}+\fft{r^4}{R^4} \rho^{\prime 2}},
\ee
where now $^{\prime}=\partial_{\phi}$. The equations of motion are
\bea\label{eom}
R^4 \fft{d}{d\phi}\left[ Af h^{-1}r^{\prime}\right] &=& 2r^3A\left( \rho^2f+ \rho^{\prime 2}
-\fft{5R^4 b^5 f}{2h^2 r^9} r^{\prime 2}\right),\nn\\
\fft{d}{d\phi} \left[ Ar^4 \rho^{\prime}\right] &=& A \left( r^4(\rho-M)+\fft{R^4 M}{h\rho^2} r^{\prime 2}\right),
\eea
where
\be
A^{-1}=\sqrt{\rho^2 r^4 f+R^4 f h^{-1} r^{\prime 2}+ r^4 \rho^{\prime 2}}.
\ee
Since the action does not depend on $\phi$ explicitly, the Beltrami identity yields
\be\label{beltrami}
A\rho^2 r^4 f=\rho_0 r_0^2 \sqrt{f_0},
\ee
where the turning point at the midsection of the string is taken to be at $\phi=0$, $r=r_0$ and $\rho=\rho_0$. This equation can be used in place of one of the equations of motion but it should still be verified that the resulting solution solves both equations of motion.

\begin{figure}[h]
\centering
\includegraphics[width=173mm]{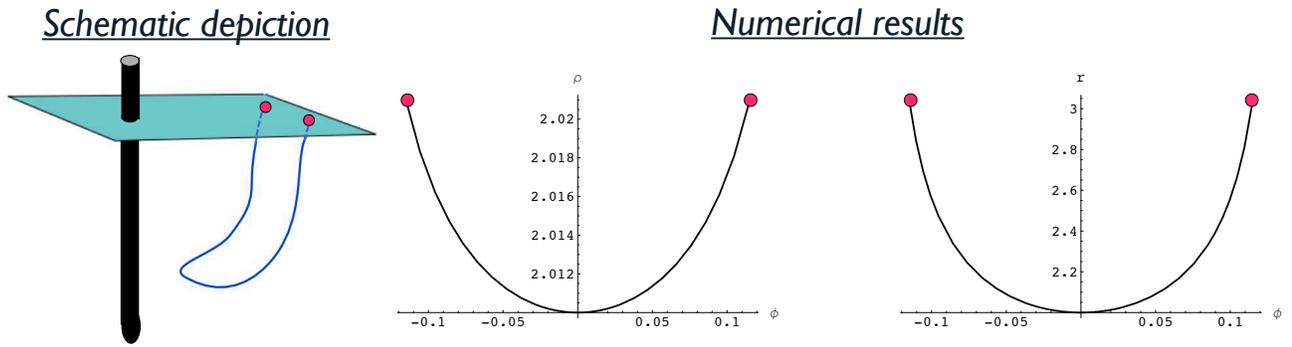}
\caption{The picture on the left is a qualitative depiction of a quark-antiquark string configuration whose endpoints are separated along the tangential direction. The center and right numerical plots show the $r$ and $\rho$ coordinates, respectively, as a function of angle $\phi$ for this type of string configuration.} 
\label{fig3}
\end{figure}

For the purpose of numerical integration, we take the boundary conditions to be at the midsection:
\be\label{mid-boundary}
r(0)=r_0,\qquad r^{\prime}(0)=0,\qquad \rho(0)=\rho_0,\qquad \rho^{\prime}(0)=0.
\ee
At this turning point, the equations of motion give
\be
r^{\prime\prime}(0)=\fft{2r_0^3\rho_0^2 h_0}{R^4}>0,\qquad \rho^{\prime\prime}(0)=\rho_0-M,
\ee
where $h_0>0$ provided that $r_0>b$. The first relation implies that the midsection of the string dips down. The second relation tells us that the midsection of the string dips towards the singularity at $\rho=0$ for $\rho_0>M$ and away from the singularity for $\rho_0<M$. A qualitative depiction of a string located outside of the horizon with only one turning point is given in the left picture of Figure \ref{fig3}, in which the bending of the string in the $\rho$ direction has been exaggerated. In reality, the bending of the string in the $\rho$ direction tends to occur at about two orders of magnitude less than the scale at which the string dips down in the $r$ direction. The center and right numerical plots of Figure \ref{fig3} show the dependence of the $r$ and $\rho$ coordinates on $\phi$ for a string with $\rho_0=2.01$. 
The behavior of the string is rather sensitive to its location within the $r-\rho$ plane. For instance, by changing the midsection turning point to be at $r_0=1$ and $\rho_0=1.99$, we find that there is a string configuration whose endpoints lie outside of the horizon and whose midsection lies inside of the horizon. The string exhibits saddlepoints on either side of its midsection where it then dips into the horizon.

If the distance between the string endpoints is increased, then at some point the energetically favorable configuration is that of two disconnected strings. This is analogous to the screening length exhibited by a quark-antiquark pair in a strongly-coupled plasma \cite{screening1,screening2}, except in this case the quark-antiquark interaction is screened by the black hole.

\section{Future Directions}

We have demonstrated that the AdS/CFT correspondence may be used to probe regions closer to the black hole singularity than expected from the four-dimensional gravity perspective. We show that quarks that interact strongly with a color-charged black hole share a number of the same features as quarks in a strongly-coupled plasma, such as quark energy loss and a quark-antiquark screening length.

There are a number of potentially useful and interesting future directions. Firstly, the stability of the Schwarzschild-black string AdS soliton still needs to be investigated. It would be nice to extend our considerations to include rotating color-charged black holes, as well as color-charged black holes in a strongly-coupled plasma. A black hole in a strongly-coupled plasma has been conjectured to be dual to a ``black droplet'' or a ``black funnel'', depending on whether the plasma interacts weakly or strongly with the black hole, respectively \cite{hubeny}. Lastly, the AdS/CFT correspondence could enable us to probe the final stages of black hole evaporation to a greater extent than has been done with the semi-classical approach in four dimensions.

\begin{acknowledgments}

We are grateful to Philip Argyres for useful discussions. This work is supported in part by NSF grant PHY-0969482 and a PSC-CUNY Award jointly funded by The Professional Staff Congress and The City University of New York.

\end{acknowledgments}

\bigskip 

\end{document}